\documentclass[12pt]{article}

\usepackage{times}
\usepackage{graphicx}
\usepackage{amsmath}
\usepackage{setspace}

\newenvironment{sciabstract}{%
\begin{quote} \bf}
{\end{quote}}

\newcounter{lastnote}
\newenvironment{scilastnote}{%
\setcounter{lastnote}{\value{enumiv}}%
\addtocounter{lastnote}{+1}%
\begin{list}%
{\arabic{lastnote}.} {\setlength{\leftmargin}{.22in}}
{\setlength{\labelsep}{.5em}}} {\end{list}}

\title{Long-lived Giant Number Fluctuations in a Swarming Granular Nematic}

\author
{Vijay Narayan,$^{1\ast}$ Sriram Ramaswamy,$^{1,2}$ Narayanan Menon$^{3}$\\
\\
\normalsize{$^{1}$CCMT, Department of Physics, Indian Institute of Science, Bangalore 560012, India}\\
\normalsize{$^{2}$CMTU, Jawaharlal Nehru Centre for Advanced Scientific Research, Bangalore 560064, India}\\
\normalsize{$^{3}$Department of Physics, University of Massachusetts, Amherst, MA 01003, USA}\\
\\
\normalsize{$^\ast$To whom correspondence should be addressed;
E-mail:  vj@physics.iisc.ernet.in} }

\date{}

\begin{document}


\baselineskip24pt


\maketitle


\begin{sciabstract}
\bf{Coherently moving flocks of birds, beasts or bacteria are
examples of living matter with spontaneous orientational order.
How do these systems differ from thermal equilibrium systems with
such liquid-crystalline order? Working with a fluidized monolayer
of macroscopic rods in the nematic liquid crystalline phase, we
find giant number fluctuations consistent with a standard
deviation growing linearly with the mean, in contrast to any
situation where the Central Limit Theorem applies. These
fluctuations are long-lived, decaying only as a logarithmic
function of time. This shows that flocking, coherent motion and
large-scale inhomogeneity can appear in a system in which
particles do not communicate except by contact.}
\end{sciabstract}


\section*{}

Density is a property that one can measure with arbitrary accuracy
for materials at thermal equilibrium simply by increasing the size
of the volume observed. This is because a region of volume $V$,
with $N$ particles on average, ordinarily shows fluctuations with
standard deviation $\Delta N \propto \sqrt{N}$, so that
fluctuations in the number density go down as $1/\sqrt{V}$.
Liquid-crystalline phases of active or self-propelled particles
(\textit{1-4}) are different, with $\Delta N$ predicted
(\textit{2-5}) to grow faster than $ \sqrt{N}$, and as fast as $N$
in some cases (\textit{5}), making density an ill-defined quantity
even in the limit of a large system. These predictions show that
flocking, coherent motion and giant density fluctuations are
intimately related consequences of the orientational order that
develops in a sufficiently dense grouping of self-driven objects
with anisotropic body shape. This has significant implications for
biological pattern formation and movement ecology (\textit {6}):
the coupling of density fluctuations to alignment of individuals
will affect populations as diverse as herds of cattle, swarms of
locusts (\textit {7}), schools of fish (\textit{8, 9}), motile
cells (\textit{10}), and filaments driven by motor proteins
(\textit{11-13}).

We report here that persistent, giant number fluctuations and the
coupling of particle currents to particle orientation arise in a
far simpler driven system, namely, an agitated monolayer of
rod-like particles shown in (\textit{14}) to exhibit
liquid-crystalline order. These fluctuations have also been
observed in computer simulations of a simple model of the flocking
of apolar particles by Chat\'{e} \textit{et al}. (\textit{15}).
The rods we use are cut to a length $\ell = 4.6 \pm 0.16$ mm from
copper wire of diameter $d = 0.8$ mm. The ends of the rods are
etched to give them the shape of a rolling-pin. The rods are
confined in a quasi-2-dimensional cell $1$ mm tall and with a
circular cross-section $13$ cm in diameter. The cell is mounted in
the horizontal plane on a permanent magnet shaker and vibrated
vertically at a frequency $f = 200$ Hz, with an amplitude,
\textsl{A}, between $0.025$ and $0.043$ mm. The resultant
dimensionless acceleration $\Gamma = (4\pi^2f^2\textsl{A})/g$,
where $g$ is the acceleration due to gravity, varies between
$\Gamma = 4$ and $\Gamma = 7$. We vary the total number of
particles in the cell, $N_{total}$ between $1500$ and $2820$.
$N_{total}$ in each instance is counted by hand. The area
fraction, $\phi$, occupied by the particles is the total projected
area of the all rods divided by the surface area of the cell.
$\phi$ varies from 35\% to 66\%. Our experimental system is
similar to those used to study the phase behaviour of inelastic
spheres (\textit{16, 17}). Galanis \textit{et al.} (\textit{18})
shook rods in a similar setup, albeit with much less confinement
in the vertical direction. The particles are imaged with a digital
camera (\textbf{Data Acquisition} (\textit{19})).

The rods gain kinetic energy through frequent collisions with the
floor and the ceiling of the cell. Since the axes of the particles
are almost always inclined to the horizontal, these collisions
impart or absorb momentum in the horizontal plane. Collisions
between particles conserve momentum, but also drive horizontal
motion by converting vertical motion into motion in the plane.
Inter-particle collisions as well as particle-wall collisions are
inelastic, and all particle motion would cease within a few
collision times if the vibrations were switched off. The momentum
of the system of rods is not conserved either, since the walls of
the cell can absorb or impart momentum. The rods are apolar --
individual particles do not have a distinct head and tail that
determine fore-aft orientation or direction of motion -- and can
form a true nematic phase. The experimental system thus has all
the physical ingredients of an active nematic (\textit{1-4}).

The system is in a very dynamic steady state, with particle motion
(see movie S1 (\textit{19})) organized in macroscopic swirls.
Swirling motions do not necessarily imply the existence of giant
number fluctuations (\textit{20}, \textit{21}), however, particle
motions in our system generate anomalously large fluctuations in
density. Figure 1A shows a typical instantaneous configuration,
and the inset to Fig. 1B shows the orientational correlation
function $G_2(r) = \langle \cos 2(\theta_i - \theta_j) \rangle$,
where $i,j$ run over pairs of particles separated by a distance
$r$ and oriented at angles $\theta_i, \, \theta_j$ with respect to
a reference axis. The angle brackets denote an average over all
such pairs and about 150 images spaced 15 seconds apart in time.
The data in the inset show that the systems with $N_{total} =
2500$ and $2820$ display quasi-long ranged nematic order, where
$G_2(r)$ decays as a power of the separation, $r$. On the other
hand, the system with $N_{total} = 1500$ shows only short-ranged
nematic order, with $G_2(r)$ decaying exponentially with $r$.
Details of the crossover between these two behaviours can be found
in (\textbf{The Shaking Amplitude $\Gamma$} (\textit{19})).
Autocorrelations of the density field as well as of the
orientation of a tagged particle decay to zero on much shorter
time-scales (\textbf{Statistical Independence of Configurations
Sampled } (\textit{19})), so we expect these images to be
statistically independent. To quantify the number fluctuations, we
extracted from each image the number of particles in subsystems of
different size, defined by windows ranging in size from
$0.1$x$0.1\,\ell^2$ to $12$x$12\,\ell^2$. From a series of images
we determined, for each subsystem size, the average $N$ and the
standard deviation, $\Delta N$, of the number of particles in the
window. For any system in which the number fluctuations obey the
conditions of the Central Limit Theorem (\textit{22}), $\Delta
N/\sqrt{N}$ should be a constant, independent of $N$. Figure 1B
shows that when the area fraction $\phi$ is large, $\Delta
N/\sqrt{N}$ is not a constant. Indeed, for big enough subsystems,
the data show giant fluctuations, $\Delta N$, in the number of
particles, growing far more rapidly than $\sqrt{N}$ and consistent
with a proportionality to $N$. For smaller average number density,
where nematic order is poorly developed, this effect disappears,
and $\Delta N/\sqrt{N}$ is independent of $N$, as in thermal
equilibrium systems. The roll-off in $\Delta N/\sqrt{N}$ at the
highest values of $N$ is a finite-size effect: for subsystems that
approach the size of the entire system, large number fluctuations
are no longer possible since the total number of particles in the
cell is held fixed.

We examine a subsystem of size $\ell$ x $\ell$ (i.e., one rod
length on a side) and obtain a time-series of particle number,
$N(t)$, by taking images at a frame rate of 300 frames/sec. From
this we determine the temporal autocorrelation, $C(t)$, of the
density fluctuations. As shown in Fig. 2, $C(t)$ decays
logarithmically in time, unlike the much more rapid $t^{-1}$ decay
of random, diffusively relaxing density fluctuations in two
dimensions. Thus the density fluctuations are not only anomalously
large in magnitude but also extremely long-lived. Indeed, these
two effects are intimately related: an intermediate step in the
theoretical argument (\textit{5}) that predicts giant number
fluctuations shows that density fluctuations at a wavenumber $q$
have a variance proportional to $q^{-2}$ and decay diffusively.
This leads to the conclusion (\textbf{Long-lived Density
Autocorrelation} (\textit{19})) that in the time regime
intermediate between the times taken for a density mode to diffuse
a particle length and the size of the system, the autocorrelation
function of the local density decays only logarithmically in time.
While the observations agree with the predicted logarithmic decay,
we cannot as yet make quantitative statements about the
coefficient of the logarithm. It is important to note that the
size of subsystem is below the scale of subsystem size at which
the standard deviation has become proportional to the mean. In
flocks and herds as well, measuring the dynamics of local density
fluctuations will yield crucial information regarding the entire
system's dynamics, and can be used to test the predictions of
Toner and Tu (\textit{2,3}).

What are the microscopic origins of the giant density
fluctuations? Both in active and in equilibrium systems, particle
motions lead to spatial variations in the nematic ordering
direction. However, in active systems alone, such bend and splay
of the orientation are predicted (\textit{5}) to select a
direction for coherent particle currents. These curvature-driven
currents in turn engender giant number fluctuations.  We find
qualitative evidence for curvature-induced currents in the flow of
particles near topological defects (\textbf{Curvature-Induced Mass
Flow} (\textit{19})). In the apolar flocking model of
(\textit{15}) particles move by hopping along their axes and then
reorienting, with a preference to align parallel to the average
orientation of particles in their neighbourhood. Requiring that
the hop be along the particle axis was sufficient to produce giant
number fluctuations in the nematic phase of the system. It was
further suggested in (\textit{15}) that the curvature-induced
currents of (\textit{5}), although not explicitly put into their
simulation, must emerge as a macroscopic consequence of the rules
imposed on microscopic motion. This suggestion is substantiated by
the work of Ahmadi \textit{et al}. (\textit{23}) who start from a
microscopic model of molecular motors moving preferentially along
biofilaments and show by coarse-graining this model that the
equation of motion for the density of filaments contains precisely
the term in (\textit{5}) responsible for curvature-induced
currents.

In our experiments we find anisotropy at the most microscopic
level of single particle motion, even at time scales shorter than
the vibration frequency $f$. In equilibrium, the mean kinetic
energy associated with the two in-plane translation degrees of
freedom of the particle are equal, by the equipartition theorem,
even if the particle shape is anisotropic. Figure 3 is a histogram
of the magnitude of particle displacements over a time
corresponding to the camera frame rate ($1/300$ sec, or
$\frac{2}{3}f^{-1}$). The displacement along and perpendicular to
the axis of the rod are displayed separately, showing that a
particle is about 2.3 times as likely to move along its length as
it is to move transverse to its length. Since the period of the
imposed vibration ($f^{-1}$) sets the scale for the mean free time
of the particles, this shows that the motion of the rods is
anisotropic even at time scales less than or comparable to the
mean free time between collisions.

We have thus presented an experimental demonstration of giant,
long-lived, number fluctuations in a 2-dimensional active nematic.
The particles in our driven system do not communicate except by
contact, have no sensing mechanisms and are not influenced by the
spatially-varying pressures and incentives of a biological
environment. This reinforces the view that in living matter as
well, simple, non-specific interactions can give rise to large
spatial inhomogeneity. Equally important, these effects offer a
counterexample to the deeply held notion that density is a
sharply-defined quantity for a large system.

\pagebreak
\begin{figure}[t]
\begin{center}
\includegraphics[width=\linewidth]{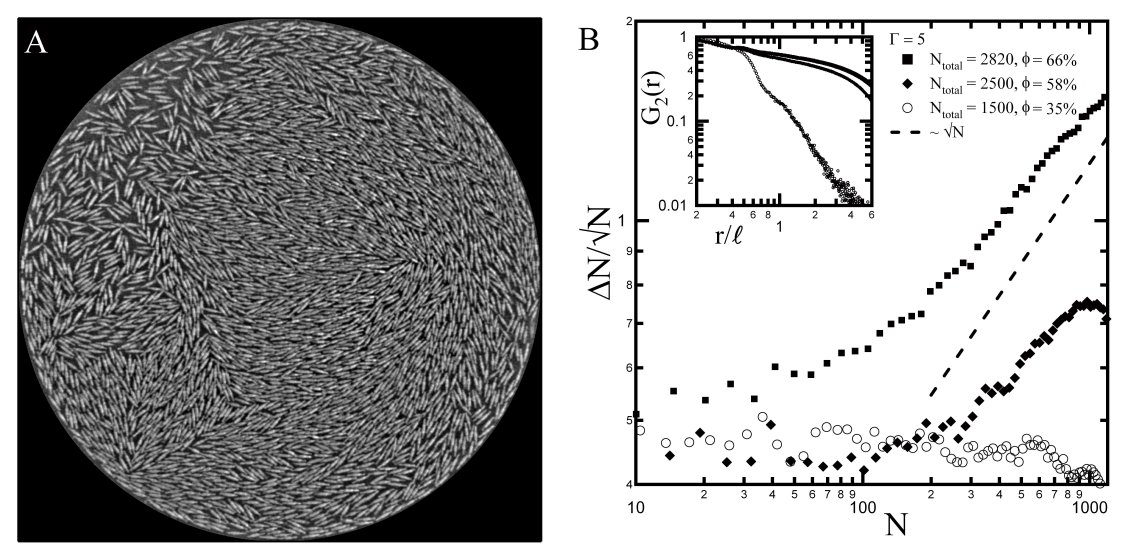}
\end{center}
\doublespacing \caption{Giant number fluctuations in active
granular rods. A) shows a snapshot of the nematic order assumed by
the rods. There are 2820 particles (counted by hand) in the cell
(area fraction ~ 66\%), being sinusoidally vibrated perpendicular
to the plane of the image, at a peak acceleration of $\Gamma = 5$.
The sparse region at the top between 10 and 11 o'clock is an
instance of a large density fluctuation. These take several
minutes to relax and form elsewhere. B) shows the magnitude of the
number-fluctuations (quantified by $\Delta N$, the standard
deviation, normalized by the square root of the mean number, $N$)
against the mean number of particles, for subsystems of various
sizes. The number fluctuations in each subsystem are determined
from images taken every 15 seconds over a period of 40 minutes
(\textbf{Methods} (\textit{19})). The squares represent the system
shown in A. It is a dense system where the nematic order is well
developed. The magnitude of the scaled number fluctuations
decreases in more dilute systems, where the nematic order is
weaker (\textbf{The Shaking Amplitude $\Gamma$} (\textit{19})).
Deviations from the Central Limit Theorem result are still visible
at an area fraction $\simeq 58 \%$ (diamonds), but not at an area
fraction $\simeq 35 \%$ (circles). The inset shows the
nematic-order correlation function as a function of spatial
separation.}\singlespacing
\end{figure}

\pagebreak

\begin{figure}[t]
\includegraphics[width=\linewidth]{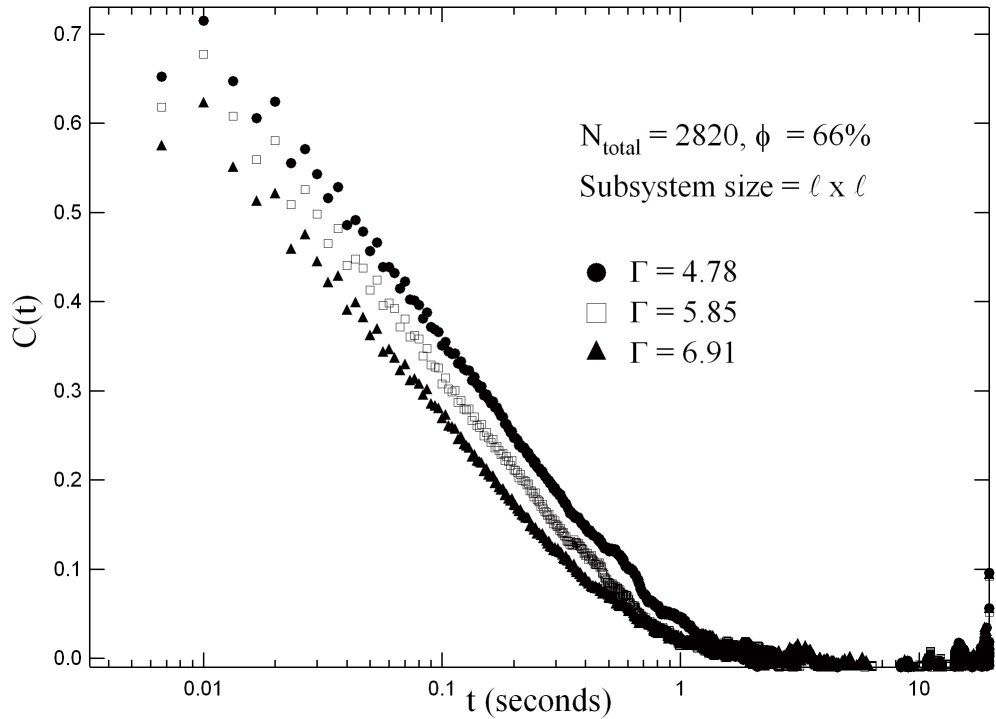}
\doublespacing \caption{The logarithmic dependence of the local
density autocorrelation $C(t) = <\varphi(0)\varphi(t)>$ (
$\varphi(t)$ is the deviation from the mean of the instantaneous
number density of particles) is a direct consequence
(\textbf{Long-lived Density Autocorrelation} (\textit{19})), and
hence a clear signature, of the large density fluctuations in the
system. It is remarkable that such a local property reflects the
dynamics of the entire system so strongly. It is seen that
increasing $\Gamma$ shortens the decay time. This is consistent
with the fact that the magnitude of the giant number fluctuations
grows with the nematic order (\textbf{The Shaking Amplitude
$\Gamma$} (\textit{19})).} \singlespacing
\end{figure}

\pagebreak

\begin{figure}[t]
\includegraphics[width=\linewidth]{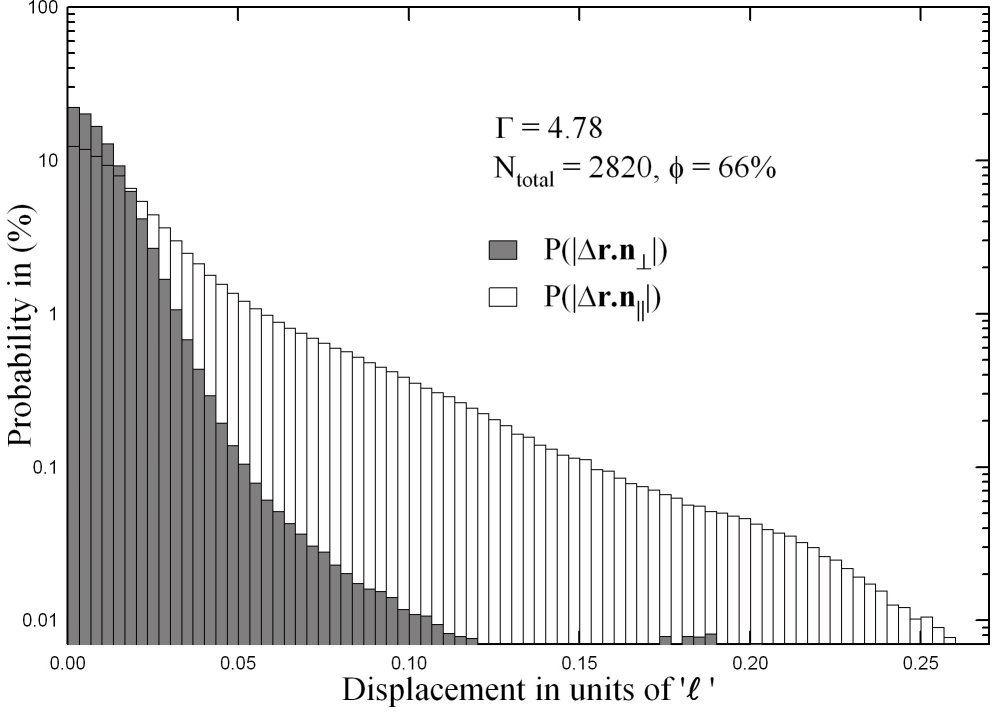}
\doublespacing\caption{The microscopic origin of the macroscopic
density fluctuations. The probability distribution of the
magnitude of the displacement along and transverse to the
particle's long axis over an interval of 1/300 of a second shows
that short time motion of the rods is anisotropic even at the
time-scale of the collision time. This anisotropy is explicitly
forbidden in equilibrium systems by the equipartition
theorem.}\singlespacing
\end{figure}

\begin{quote}
{\bf References and Notes}

\begin{enumerate}

\item T. Vicsek, A. Czirok, E. Ben-Jacob, I. Cohen, O. Shochet,
\textit{Phys. Rev. Lett.} \textbf{75}, 1226 (1995).

\item J. Toner, Y. Tu, \textit{Phys. Rev. Lett.} \textbf{75}, 4326
(1995).

\item J. Toner, Y. Tu, \textit{Phys. Rev. E} \textbf{58}, 4828
(1998).

\item J. Toner, Y. Tu, S. Ramaswamy, \textit{Ann. Phys.}
\textbf{318}, 170 (2005).

\item S. Ramaswamy, R. A. Simha, J. Toner, \textit{Europhys.
Lett.} \textbf{62}, 196 (2003).

\item C. Holden, \textit{Science} \textbf{313}, 779 (2006).

\item J. Buhl \textit{et al.}, \textit{Science} \textbf{312}, 1402
(2006).

\item N. Makris \textit{et al.}, \textit{Science} \textbf{311},
660 (2006).

\item C. Becco \textit {et al.}, \textit {Physica A},
\textbf{367}, 487 (2006).

\item B. Szab\'{o} \textit{et al.}, \textit{Phys. Rev. E}
\textbf{74}, 061908 (2006).

\item F. N\'ed\'elec, T. Surrey, A. C. Maggs , S. Leibler,
\textit{Nature} \textbf{389}, 305 (1997).

\item D. Bray, \textit{Cell Movements: From Molecules to Motility}
(Garland Publishing, New York, 2001).

\item H. Gruler, U. Dewald, M. Eberhardt, \textit{Eur. Phys. J. B
}\textbf{11}, 187 (1999).

\item V. Narayan, N. Menon, S. Ramaswamy, \textit{J. Stat. Mech.}
(2006) P01005

\item H. Chat\'{e}, F. Ginelli, R. Montagne, \textit{Phys. Rev.
Lett.} \textbf{96}, 180602 (2006).

\item A. Prevost, D. A. Egolf, J. S. Urbach, \textit{Phys. Rev.
Lett.} \textbf{89}, 084301 (2002).

\item P. M. Reis, R. A. Ingale, M. D. Shattuck, \textit{Phys. Rev.
Lett.} \textbf{96}, 258001 (2006).

\item J. Galanis, D. Harries, D. L. Sackett, W. Losert, R. Nossal,
\textit{Phys. Rev. Lett.} \textbf{96}, 028002 (2006).

\item Supporting Online Material \\www.sciencemag.org \\Materials
and Methods\\SOM Text\\ Figs. S1, S2, S3, S4, S5, S6 \\ Movie S1,
S2

\item D. L. Blair, T. Neicu, A. Kudrolli, \textit{Phys. Rev. E}
\textbf{67}, 031303 (2003).

\item I. Aranson, D. Volfson, L. S. Tsimring, \textit{Phys. Rev.
E} \textbf{75}, 051301 (2007).

\item W. Feller, \textit{An introduction to Probability Theory and
its Applications, Volume I}, (John Wiley \& Sons, New York, 3rd
edition, 2000)

\item A. Ahmadi, T. B. Liverpool, M. C. Marchetti, \textit{Phys.
Rev. E} \textbf{74}, 061913 (2006).

\end{enumerate}
\end{quote}


\begin{scilastnote}
\item \textbf{Acknowledgements.} We thank V. Kumaran, P. Nott and
A. K. Raychaudhuri for generously letting us use their
experimental facilities. VN thanks Sohini Kar for help with some
of the experiments. VN and SR respectively thank the Council for
Scientific and Industrial Research, India and the Indo-French
Centre for the Promotion of Advanced Research (grant 3504-2) for
support. The Centre for Condensed Matter Theory is supported by
the Department of Science and Technology, India. NM acknowledges
financial support from the National Science Foundation under
NSF-DMR 0606216 and 0305396.
\end{scilastnote}


\clearpage

\doublespacing

\begin{center}
\Large{Supporting Online Material}
\end{center}

\section*{Methods}

\subsection*{Data Acquisition}

To study the instantaneous spatial configurations of the system,
we used a digital camera (Canon PowerShot G5) at a resolution of
2592x1944 pixels. Images of the entire cell were taken at roughly
15-second intervals from which we extracted the position and
orientation of all particles (details of the image analysis are
given in the section \textbf{Image Analysis}). The degree of
orientational order was quantified by the nematic order
correlation function
\begin{equation*}
G_2(r) = <\Sigma_{i,j}cos[2(\theta_i - \theta_j)]>
\end{equation*}
Here the labels $i,\,j$ run over all pairs of particles with
separation $r$, and the angular brackets represent an average over
the images. Movie S1 was made from images taken using this camera
(see section \textbf{Description of the Movie material}).

In order to probe the dynamics of the system, we used a high speed
camera (Phantom v7) to track particles in a  region of the entire
sample consisting of 400 - 500 particles. Images were taken at 300
frames per second for a period of 20 seconds and at a resolution
of 800x600 pixels. Movie S2 was made using this camera (see
section \textbf{Description of the Movie material}).

\subsection*{Image Analysis}
\label{IA}

The centre-of-mass coordinates and orientation of all the
particles were extracted from each image by image analysis
routines. Particle identification in the image analysis was
complicated by the high number density of particles and the
consequent overlaps in particle footprints. Furthermore, the
particles are of unusual shape and not perfectly monodisperse. The
image analysis was done using the ImageJ program. To separate
particles that were in contact, the contrast at the edges of the
particles was enhanced, followed by a sequence of erosion and
opening operations. Typically 90\% of the particles were
identified using this procedure. Two further special cases had to
be dealt with for the remaining particles. The first was that
erosion often caused the narrow ends of the particles  to detach,
yielding spurious identification as individual particles. To
eliminate them, we imposed both a minimum-area threshold as well
as a condition that eliminated the smaller of a pair of particles
if they were lying too close in the longitudinal direction. The
second class of problem came from particle footprints whose tips
fused together forming a composite particle along their length.
These cases were easy to identify both by their large area and
their long aspect ratio. We resolved such cases by assigning to
the composite object, two particles lying symmetrically about the
original position along the original axis. They were assigned the
same orientation as the originally identified object.

When analysing data to study statics, the focus was on identifying
each particle in the system. Since we were trying to quantify the
fluctuations in the number density, it was imperative that no
spurious number fluctuations arose from the image analysis. Both
direct visual verification of the accuracy of the identification
and the roll-off in Fig. 1B assure us that the fluctuations due to
errors in the image analysis are much smaller than the inherent
number fluctuations in the system.

When studying the dynamics of the particles, we developed a
standard particle-tracking code based on a minimum distance
criterion between particles in successive frames. This approach
was effective because particles typically moved only about a
fiftieth of their length between successive images. If no particle
was found in a later frame that uniquely corresponded to the
initial one, the particle location was held for comparison to
later images, so that a particle ``lost'' in a particular frame
might be recovered later.

\section*{Curvature-Induced Mass Flow}
\label{CIV}

In reference (S1) the authors argue that curvature in the nematic
orientation gives rise to mass motion of particles, and that the
great abundance and slow decay of large-scale curvature leads
ultimately to giant number fluctuations. To get direct evidence of
this mechanism in our experiment would require mapping out the
curvature field and correlating it to the local mass flux. For
simplicity, we look instead for individual instances of large
curvature and the corresponding motion. This is conveniently done
by focusing on the regions around topological defects in an
otherwise well-oriented nematic. Two types of defects, strength
+1/2 and -1/2, are seen with their associated nematic director
fields shown in Fig. S1. A topological defect is a localized
imperfection in the state of order which cannot be removed by
continuous deformations of the order parameter. For a more
detailed understanding of defects in nematics see reference (S2).
What is crucial for our discussion is that the orientation field
around the +1/2 defect in figure B is polar and breaks the
fore-aft symmetry, while that around the -1/2 defects is
three-fold symmetric. A curvature-induced drift mechanism would
result in particle motion in the directions indicated by the
arrows in figures A and B. The symmetry of the field around the
strength -1/2 defect will result in no net motion, while the
curvature around the +1/2 defect has a well-defined polarity and
hence should move in the direction of its ``nose'' as shown in the
figure. Movie S2, stills from which are seen in Fig. S1, shows
precisely this behaviour.

\section*{The Shaking Amplitude $\Gamma$}

$\Gamma$, the dimensionless acceleration, is one of the 2
quantities we vary to explore our parameter space, the other being
density. Na\"{\i}vely one would expect that $\Gamma$ in our
nonequilibrium phase diagram is analogous to the temperature in
thermal equilibrium systems -- the harder one shakes the system,
the more disordered it gets, making a direct correspondence with
the temperature. We observe a more complicated, non-monotonic
trend.

Figure S2 shows the nematic order in the system as a function of
density and $\Gamma$. Increasing density clearly increases nematic
order. At the three lowest densities, increasing $\Gamma$
increases the nematic order. At $N_{total} = 2820$ however,
increasing $\Gamma$ decreases the nematic order. The crossover in
the behaviour occurs at $N_{total}=2500$ where we see that
$G_2(r)$ for $\Gamma = 5$ and $\Gamma = 6$ are nearly
indistinguishable. Put together, this suggests that these points
are near a re-entrant portion of the isotropic-nematic transition.

The complicated role $\Gamma$ plays in the system's dynamics is
apparent even in the giant number fluctuations. Figure S3 shows
$\Delta N / \sqrt{N}$ as a function of $N$ for 4 different
densities. We see that there are two regimes of fluctuations, most
clearly distinguishable in the $N_{total} = 2500$ curve: (i)
$\Delta N \propto \sqrt{N}$ at small $N$, and (ii) $\Delta N
\propto N$, the giant fluctuations, at large $N$. The $\sqrt{N}$
at smaller $N$ arises from uncorrelated short-time diffusion of
particles and exists even in the isotropic phase. The giant
fluctuations dominate, as expected from (S1), at large $N$, and
should be larger in systems with stronger nematic order. And
indeed, increasing density (and hence the nematic order)
suppresses the $\sqrt{N}$-fluctuations and enhances the
$N$-fluctuations, as is seen in the $N_{total}=2000$ and
$N_{total}=2500$ curves in Fig. S3. At lower subsystem sizes the
$\sqrt{N}$-fluctuations will dominate. Since these are stronger
for dilute systems, the magnitude of fluctuations is greater for
$N = 2000$. For large enough subsystems, the contribution scaling
as $N$, peculiar to the active nematic, will overwhelm the
component that grows as $\sqrt{N}$. The $N_{total}=2500$ curve
shows a sharper rise than the $N_{total}=2000$ curve and
ultimately the fluctuations in the former outgrow those in the
latter system.

To verify whether increased nematic order does indeed increase the
magnitude of fluctuations, we plot $\Delta N / \sqrt{N}$ as a
function of $N$ for 2 different values of $\Gamma$ in Fig. S4. As
expected in the dilute systems, where $\Gamma$ increases the
nematic order, it increases the magnitude of fluctuations, whereas
for $N_{total}=2820$, increasing $\Gamma$ reduces the magnitude of
fluctuations. This suggests that $\Gamma$ sets the scale of two
different stochastic processes at the level of an individual rod
-- fore-aft motion, which promotes nematic order, and rotational
motion, which disrupts it.

\section*{Long-lived Density Autocorrelation}

The principal result described in (S1) is that density modulations
at long-wavelength $\lambda = 2 \pi / q$ in active nematics have a
variance proportional to $q^{-2}$, i.e., at larger and larger
length scales, the system shows larger and larger statistical
inhomogeneities. Defining $\varphi({\bf r},t) \equiv \rho({\bf
r},t) - \rho_0$ where $\rho({\bf r},t)$ is the density at ${\bf
r}$ at time $t$ and $\rho_0$ its time average, the prediction is
that the static structure factor $S(\textbf{q}) \equiv
(1/\rho_0)\int_{\bf r} \mbox{e}^{i{\bf q} \cdot {\bf r}} \langle
\varphi({\bf 0}) \varphi({\bf r}) \rangle \propto 1/q^2$.
Converting to real space, the standard deviation $\Delta N$ would
therefore grow as $N^{\frac{1}{2} + \frac{1}{d}}$ in $d$ space
dimensions. In 2 dimensions, thus, $\Delta N \propto N$:
fluctuations grow linearly with the mean.

\subsection*{Dynamics}
\label{Dynamics}

This global result has a sharp signature in the dynamics and
statistics of the number of particles in a local region, as
measured by the local density autocorrelation function
\begin{equation}
\label{C_t} C(t) = \langle \varphi({\bf x},0)\varphi({\bf x},t)
\rangle.
\end{equation}
For independently diffusing Brownian particles
\begin{equation}
\label{C_t_B}C(t) = \int {\mbox{d}^d q \over (2 \pi)^d} e^{-q^2Dt}
\sim t^{-d/2}
\end{equation}
so that the local density autocorrelation function for a
two-dimensional thermal equilibrium system of non-interacting
Brownian particles decays as $t^{-1}$. This holds even for a
diffusing density field coupled to the director field in a nematic
liquid crystal at thermal equilibrium. For an active nematic,
however, as shown in (S1), Eq. (\ref{C_t_B}) has to be modified to
account for the anomalous $q^{-2}$ abundance of the Fourier modes
of the density at small wavenumber:
\begin{equation}
\label{C_t_A}C(t) \sim \int^{2\pi/a}_{2\pi/L} \mbox{d}q
q^{d-1}\frac{e^{-q^2Dt}}{q^2}
\end{equation}
with limits of integration determined by the coarse-graining scale
$a$ that we use to define the density, and the system-size $L$.
This implies that for timescales between those corresponding to
$a$ and to $L$, $C(t)$ should decay as $\ln (L^2/Dt)$. Our
experimental results are consistent with such a decay. We also
find that coefficient of this logarithm depends on the
coarse-graining scale $a$ (Fig. S5).

\section*{The Statistical Independence of Configurations Sampled}

To verify whether the images taken at 15-second intervals were
indeed statistically independent, we calculated two quantities: 1)
the local density autocorrelation $C(t)$ (discussed in section
\textbf{Dynamics}) and 2) the autocorrelation of the orientation
of a tagged particle, $C_{\theta}(t) = <cos[2\{\theta(0)-
\theta(t)\}]>$. The angular brackets denote an average over all
particles. Figure S6 shows that after an initially slow decay, for
about a tenth of a second, $C_{\theta}$ decays logarithmically to
zero as in two-dimensional thermal equilibrium nematics. This
decay is over about 10 seconds. Thus, both $C_{\theta}(t)$ and
$C(t)$ decay fast enough for us to be certain that the
configurations we are sampling are statistically independent.

\section*{Description of the Movie Material}
\label{movies}

Movie S1 is a recording of a typical experimental run. The system
is characterised by large inhomogeneities and large mass-fluxes.
The system contains 2,820 ($\phi \simeq 66\%$) particles and is
being agitated at $\Gamma = 5$. The time is indicated at the
bottom-left in minutes:seconds.

Movie S2 zooms into a small region of the sample. In particular,
it compares the behaviour of $+1/2$ and $-1/2$ defects. A strength
$+1/2$ defect is seen to move across the field of view of the
camera starting at the bottom-left and the strength $-1/2$ defect
is seen to remain relatively stationary as is argued in the
section \textbf{Curvature-Induced Mass Flow}. The time is
indicated in seconds at the bottom-left.

\pagebreak

\begin{figure}[t]
\includegraphics[width=\linewidth]{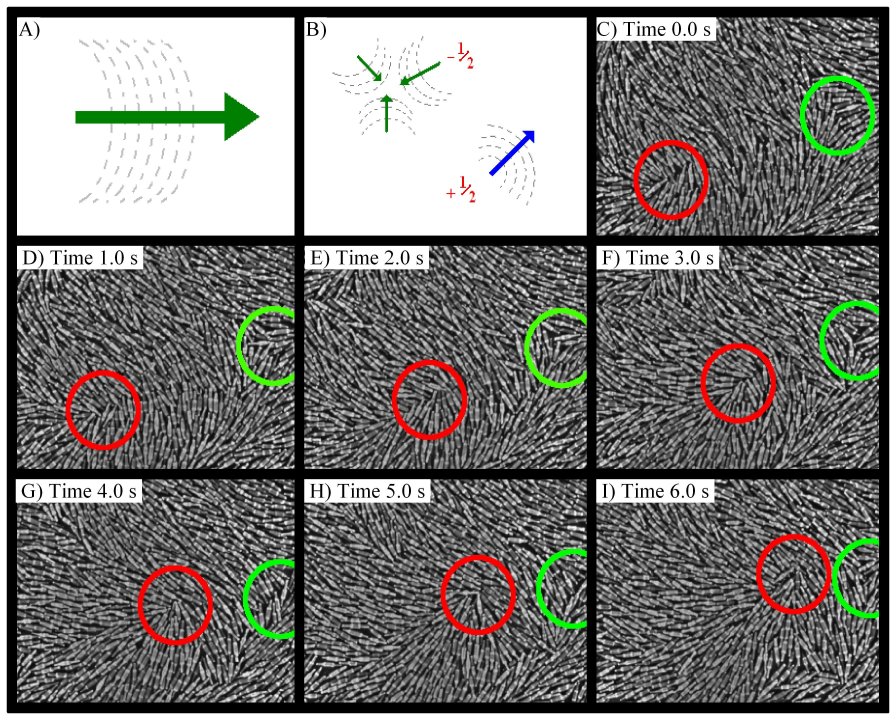}
\hfill \label{Image_CIV}
\end{figure}

\noindent\textbf{Figure S1}. Schematic image A) shows the the
motion predicted (S1) to arise in response to a curvature in the
nematic director field. Fig B) applies this idea to the director
fields around topological defects of strength +1/2 and -1/2.
Particle orientations are shown in grey and the resulting particle
currents are in the direction of the arrow. In B) the three
competing particle streams would hold a -1/2 defect stationary
whereas the +1/2 would move along the blue arrow. The time-lapse
images C) through I) show exactly this behaviour in our
experiment. The strength -1/2 defect (in the green circle) moves
very little, while the strength +1/2 defect (encircled in red)
moves substantially and systematically along its nose.

\pagebreak

\begin{figure}[t]
\includegraphics[width=\linewidth]{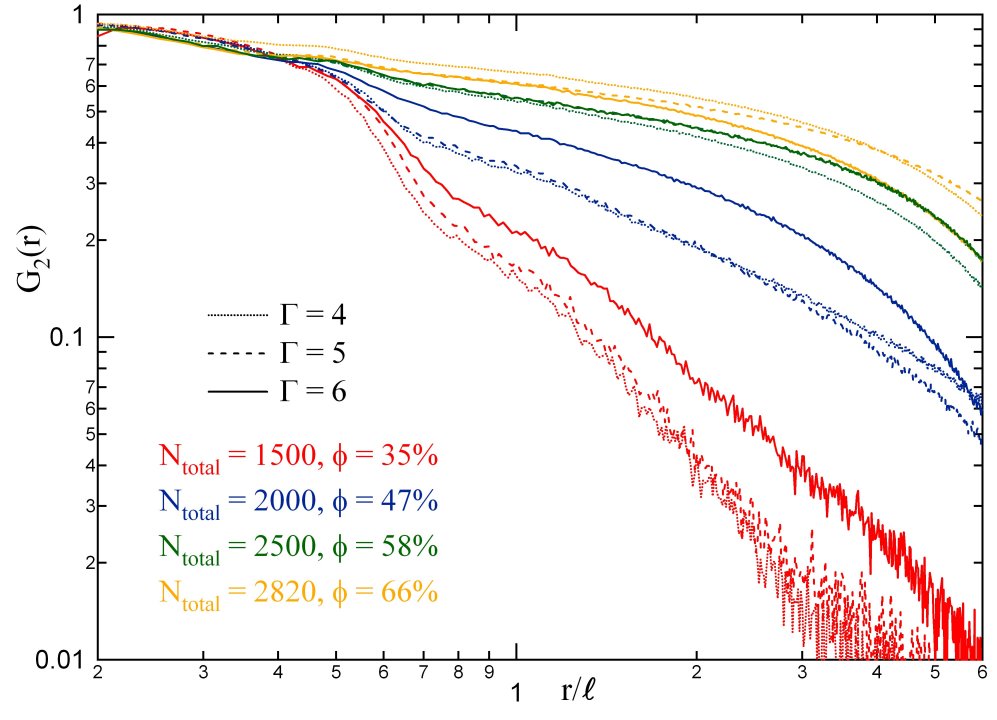}
\hfill \label{g2r}
\end{figure}

\noindent\textbf{Figure S2}. The nematic correlation function
$G_2(r)$ for various values of $\Gamma$ and density. The curves
for $N_{total}=2500$ at $\Gamma$ = $5$ and $6$ cannot be
distinguished at our resolution and, we believe, pass close to the
isotropic-nematic transition.

\pagebreak

\begin{figure}[t]
\includegraphics[width=\linewidth]{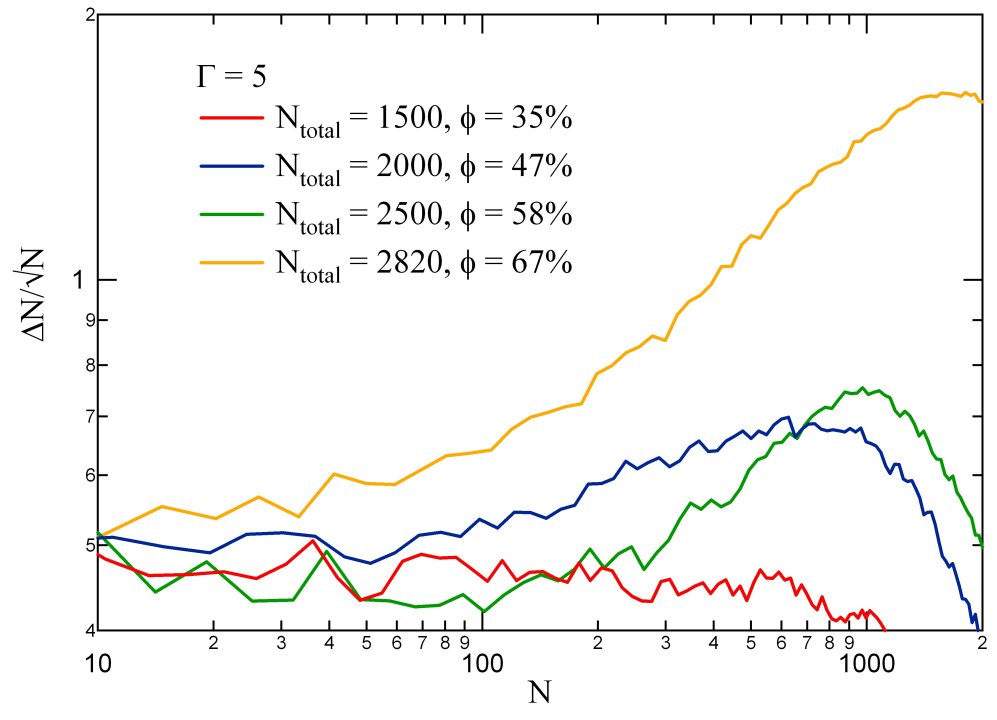}
\hfill \label{GNF}
\end{figure}

\noindent\textbf{Figure S3}. Number fluctuations at different
densities: the magnitude of the density fluctuations grows with
increasing nematic order.

\pagebreak

\begin{figure}[h]
\includegraphics[width=\linewidth]{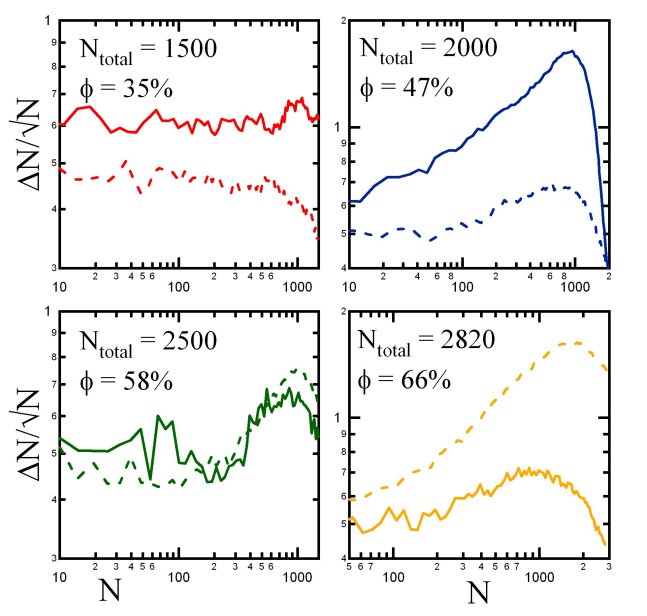}
\hfill \label{GNF_gamma}
\end{figure}

\noindent\textbf{Figure S4}. The dependence of the magnitude of
relative density fluctuations on $\Gamma$. The dashed curves
represent $\Gamma = 5$ and the solid curves $\Gamma = 6$.

\pagebreak

\begin{figure}[t]
\includegraphics[width=\linewidth]{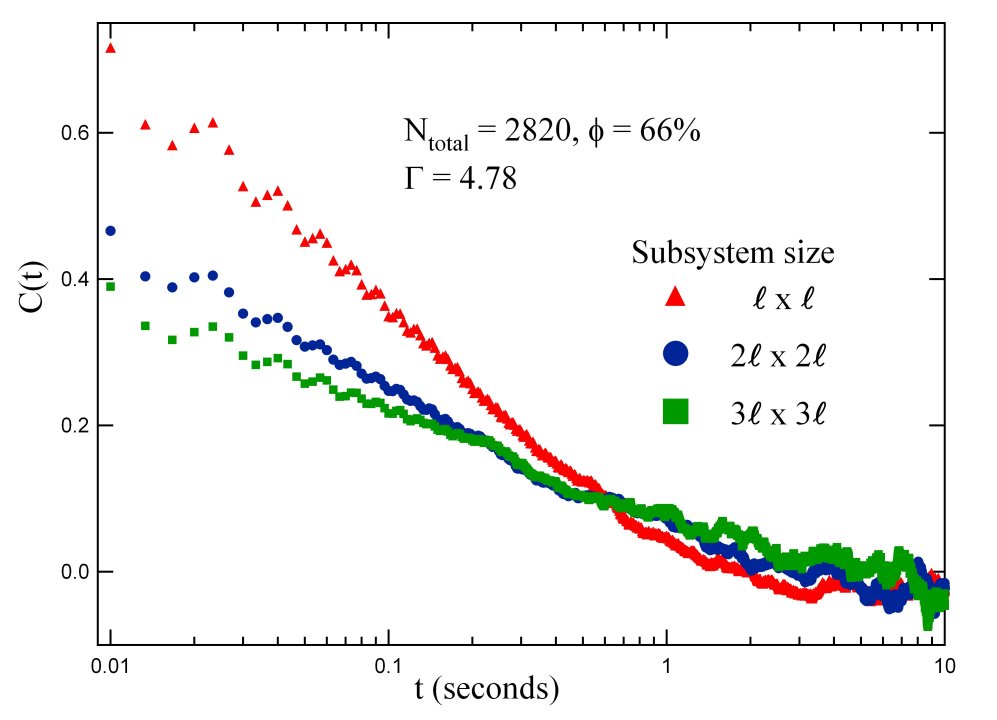}
\label{Image_C_t}
\end{figure}

\noindent\textbf{Figure S5}. The figure shows $C(t)$ evaluated
using for three different subsystems of linear size, $a$. The data
for the two larger subsystem sizes has been scaled up by the
subsystem area to facilitate viewing.

\pagebreak

\begin{figure}[t]
\includegraphics[width=\linewidth]{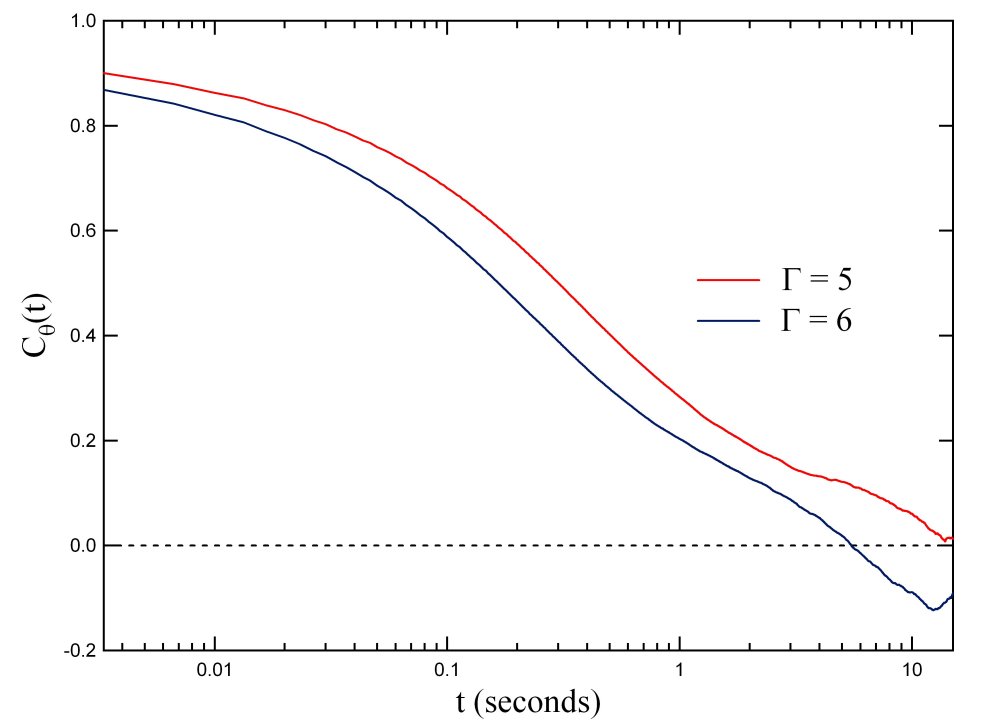}
\hfill \label{Image_C_theta}
\end{figure}

\noindent \textbf{Figure S6}. The autocorrelation of the
orientation of a tagged particle, $C_{\theta}(t) = \langle
cos[2\{\theta_i(\tau) - \theta_i(\tau+t)\}]\rangle_{i, \tau}$
where $i$ runs over all particles.\\

\section*{References}

\noindent \textbf{S1}. S. Ramaswamy, R. A. Simha, J. Toner,
\textit{Europhys. Lett.} \textbf{62}, 196 (2003).

\noindent \textbf{S2}. P. G. de Gennes, J. Prost, \textit{The
Physics of Liquid Crystals, Clarendon}, (Clarendon, Oxford 1993);
chap. 4, p. 169

\end{document}